\documentstyle[aps,prl,epsfig,twocolumn]{revtex}

\def\be{\begin{equation}}
\def\ee{\end{equation}}
\def\bea{\begin{eqnarray}}
\def\eea{\end{eqnarray}}
\def\bma{\begin{mathletters}}
\def\ema{\end{mathletters}}
\def\C{\hbox{$\mit I$\kern-.7em$\mit C$}}

\begin{document}
\draft

\title{Separability and distillability of multiparticle quantum systems}

\author{W. D\"ur$^1$, J. I. Cirac$^1$, and R. Tarrach$^2$}

\address{$^1$Institut f\"ur Theoretische Physik, Universit\"at Innsbruck,
A-6020 Innsbruck, Austria}
\address{$^2$ Departament d'Estructura i Constituents de la Mat\`eria,
Universitat de Barcelona, Spain}

\date{\today}

\maketitle

\begin{abstract}
We present a family of 3--qubit states to which any arbitrary state
can be depolarized. We fully classify those states with respect to their
separability and distillability properties. This provides a sufficient
condition for nonseparability and distillability for arbitrary states.
We generalize our results to $N$--particle states.
\end{abstract}


\narrowtext


Entanglement is an essential ingredient in most applications of Quantum
Information. It arises when the state of a multiparticle system is
non--separable; that is, when it cannot be prepared locally by acting on
the particles individually. Although in recent years there have been
important steps towards the understanding of this feature of Quantum
Mechanics, we do not know yet how to classify and quantify entanglement.

Several years ago, entanglement was thought to be directly connected to
the violation of Bell--type inequalities \cite{Be65}. However, Werner
\cite{We89} introduced a family of mixed states describing a pair of
two--level systems (qubits), the so--called Werner states (WS) which,
despite of being non--separable, do not violate any of those
inequalities \cite{Po94}. This family is characterized by a single
parameter, the fidelity $F$, which measures the overlap of WS with a
Bell (maximally entangled) state. A WS with $F>1/2$ is non--separable,
whereas if $F\le 1/2$ it is separable. WS have played an essential role
in our understanding of the quantum properties of two--qubit states
\cite{Wo98}. First of all, any state of two qubits can be reduced to a
WS by acting locally on each qubit (the so--called depolarization
process) \cite{Be96}. This automatically provides a sufficient criterion
to determine if a given state is nonseparable \cite{Pe96,Ho97}. On
the other hand, Bennett {\it et al.} \cite{Be96} showed how one can
obtain WS of arbitrarily high fidelity out of many pairs with $F>1/2$ by
using local operations and classical communication. This process, called
distillation (or purification), is one of the most important concepts in
quantum information theory. When combined with teleportation
\cite{Be93}, it allows one to convey secret information via quantum privacy
amplification \cite{De96} or to send quantum information over noisy
channels \cite{Be93,Noisy}. 
In the case of two qubits, the partial transposition\cite{PT} turned out to be a central 
tool in the classification of such systems, providing necessary and sufficient 
conditions for separability\cite{Pe96,Ho97} and distillability\cite{Ho97b}.

The description of the entanglement and distillability
properties of systems with more than two particles
is still almost unexplored (see Refs. \cite{Mu98,multi},
however). In this Letter we provide a complete classification of
a family of states of three--particle systems. These states are
characterized in terms of four parameters and play the role of
WS in such systems. In order to classify 3--qubit states with
respect to their entanglement, we define five different classes.
To display the distillability properties, we introduce a
powerful purification procedure. We also generalize our results
to systems of $N$ qubits. Among other things, this allows us to
give the necessary and sufficient separability and
distillability conditions of mixtures of a maximally entangled
state and the completely depolarized state \cite{Br99,Vi99}.
Moreover, since all states can be depolarized to the form we
analyze, our results automatically translate into sufficient
conditions for nonseparability and distillability of general
multiparticle systems. This paper is organized as follows: first
we give a classification of entangled states of three qubits;
then we analyze the separability and distillability properties
of a certain class of three--qubit states, and then we
generalize our results to more than three qubits.
\\
{\bf 1. Classification of states}

Let us consider three qubits $A$, $B$, and $C$. We classify their
possible states according to whether they are separable or not with
respect to the different qubits. In particular, according to whether
they can be written in one or more of the following forms:
\bma
\label{all}
\bea
\label{ABC}
\rho &=& \sum_i |a_i\rangle_A\langle a_i| \otimes |b_i\rangle_B\langle b_i|
 \otimes |c_i\rangle_C\langle c_i|\\
\label{A(BC)}
\rho &=& \sum_i |a_i\rangle_A\langle a_i| \otimes |\varphi_i\rangle_{BC}\langle \varphi_i|\\
\label{B(AC)}
\rho &=& \sum_i |b_i\rangle_B\langle b_i| \otimes |\varphi_i\rangle_{AC}\langle \varphi_i|\\
\label{C(AB)}
\rho &=& \sum_i |c_i\rangle_C\langle c_i| \otimes |\varphi_i\rangle_{AB}\langle \varphi_i|.
\eea
\ema
Here, $|a_i\rangle$, $|b_i\rangle$ and $|c_i\rangle$ are (unnormalized)
states of systems $A$, $B$ and $C$, respectively, and $|\varphi_i\rangle$ are
states of two systems. We have the following complete set of disjoint
classes of states:

{\bf Class 1} {\it Fully inseparable states}: Those are states that
cannot be written in any of the above forms (\ref{all}). An example is
the GHZ state \cite{Gr89} $|\Psi^+_0\rangle = (|000\rangle +
|111\rangle)/\sqrt{2}$ \cite{note2}, which is a maximally entangled
state of the three qubits.

{\bf Class 2} {\em 1-qubit biseparable states:} biseparable states with
respect to qubit $A$ are states that are separable with respect to the
first qubit, but non--separable with respect to the other two. That is,
states that can be written in the form (\ref{A(BC)}) but not as
(\ref{B(AC)}) or (\ref{C(AB)}). A trivial example would be a state
$|0\rangle_A\otimes|\Phi^+\rangle_{BC}$, where
$|\Phi^+\rangle=(|00\rangle+|11\rangle)/\sqrt{2}$ is a maximally
entangled state of two qubits.

{\bf Class 3} {\it 2-qubit biseparable states}: biseparable states with
respect to qubits $A$ and $B$ are states that are separable with respect
to the first qubit and second qubit, but non--separable with respect to
the third one. That is, states that can be written in the forms
(\ref{A(BC)}) and (\ref{B(AC)}) but not as (\ref{C(AB)}). For
examples, see below.

{\bf Class 4} {\it 3-qubit biseparable states}: Those are states that
can be written as (\ref{A(BC)}), (\ref{B(AC)}), and (\ref{C(AB)}),
but not as (\ref{ABC}). For an example, see Ref.\ \cite{Be98}.

{\bf Class 5} {\em Fully separable states}: these are states that can be
written in the form (\ref{ABC}). A trivial example is a product state
$|0\rangle_A\otimes|0\rangle_B\otimes|0\rangle_C$.

One can also consider the process of distillation of multiparticle
states and relate it to this classification. Assume that we have many
trios (3 qubits) in the same state $\rho$ and we can only perform local
operations. Then, if the state $\rho$ belongs to the class $k$ it is
clear that it is not possible to obtain one trio of a class $k'<k$ (for
instance, one cannot convert a 2--qubit biseparable state into a
1--qubit biseparable state). In some cases, however, one may produce
some maximally entangled states within one class: (i) Within the fully
inseparable states, one may be able to distill a GHZ state; (ii) Within
the 1--qubit biseparable states, one may distill a $|\Phi^+\rangle$
state of two particles. The specific conditions under which this is
possible are not known. On the other hand, if one has some extra
entanglement, one may {\it activate} some of the states and change the
corresponding class. For example, if we have a biseparable state with
respect to particles $A$ and $B$ (class 3) whose density operator has a
negative partial transpose \cite{PT} with respect to $C$ and we have some extra
states $|\Phi^+\rangle_{AB}$ at our disposal, then one can distill a GHZ
state (class 1)\cite{note}. This leads to the interesting result that
even though particle $A$ is disentangled from $BC$ and $B$ from $AC$,
with entanglement between $AB$ we can obtain a fully inseparable state.
Note also that this way of activating some hidden entanglement is
different from the one presented in Ref.\ \cite{Ho99}. It is also worth
mentioning that it is not known whether the entanglement of
the states in Class 4 can be activated in any form.
\\
{\bf 2. Three--qubit systems}

Let us define the orthonormal GHZ--basis \cite{Gr89}
\be
|\Psi^\pm_j\rangle \equiv \frac{1}{\sqrt{2}} (|j\rangle_{AB}|0\rangle_C
  \pm |(3-j)\rangle_{AB}|1\rangle_C), \label{GHZbasis}
\ee
where $|j\rangle_{AB}\equiv |j_1\rangle_A|j_2\rangle_B$ with
$j=j_1j_2$ in binary notation. For example,
$|\Psi_0^{\pm}\rangle=\frac{1}{\sqrt{2}}(|000\rangle \pm
|111\rangle)$ are standard GHZ states. We consider a family of
three--qubit states of the form
\be
\label{rho0}
\rho_3 = \sum_{\sigma=\pm} \lambda_0^\sigma |\Psi^\sigma_0\rangle\langle
  \Psi^\sigma_0| + \sum_{j=1}^3 \lambda_j (|\Psi^+_j\rangle\langle \Psi^+_j|
  + |\Psi^-_j\rangle\langle \Psi^-_j|).
\ee
The $\lambda$s are positive numbers and are restricted by
tr$(\rho_3)=1$, and therefore the states are characterized by
four parameters. We will assume that the labelling has been
chosen so that $\Delta\equiv \lambda_0^+- \lambda^-_0 \ge 0$. By
using random local operations one can convert any state to this
form while keeping the values of $\lambda_0^\pm\equiv \langle
\Psi^\pm_0|\rho|\Psi^\pm_0\rangle$ and $2\lambda_j \equiv \langle
\Psi^+_j|\rho|\Psi^+_j\rangle +
\langle \Psi^-_j|\rho|\Psi^-_j\rangle$ unchanged
\cite{Depol}. Thus, any state can be reduced to this form using
this depolarization procedure. Note that one has the freedom to
choose a local basis $\{|0\rangle,|1\rangle\}$ in A,B and C. In
this sense, the state $|\Psi_0^+\rangle$ is an arbitrary
maximally entangled state.

We will need later on the conditions under which the operator
$\rho_3$ has negative partial transpose\cite{PT} with respect to
each qubit. One can readily check that
\bea
\label{pt}
\rho_3^{T_A} \ge 0 \quad {\rm iff} \ \Delta \le 2 \lambda_{2} \nonumber \\
\rho_3^{T_B} \ge 0 \quad {\rm iff} \ \Delta \le 2 \lambda_{1} \\
\rho_3^{T_C} \ge 0 \quad {\rm iff} \ \Delta \le 2 \lambda_{3}.\nonumber
\eea
\\
{\bf 2.1. Separability}

We start out by analyzing the separability properties of the
states (\ref{rho0}). We show: (i) $\rho_3$ can be written in the
form (\ref{A(BC)}) iff $\rho_3^{T_A}\ge 0$ [and analogously for
(\ref{B(AC)}) and (\ref{C(AB)}) with $\rho_3^{T_B}\ge 0$ and
$\rho_3^{T_C}\ge 0$, respectively]. (ii) $\rho_3$ can be written
as (\ref{ABC}) iff $\rho_3^{T_A},\rho_3^{T_B},\rho_3^{T_C} \ge
0$. These results give rise to the classification of the states
given in Table I.

(i) If $\rho_3^{T_A}\ge 0$ then $\rho_3$ can be written in the
form (\ref{A(BC)}) [the opposite is true given the fact that
positive partial transposition is a necessary condition for
separability \cite{Pe96}]. The idea of our proof is to define an
operator $\tilde\rho$ which can be written as (\ref{A(BC)}) and
can be brought into the form (\ref{rho0}) by local operations,
which is sufficient to show the separability of $\rho_3$ since a
separable operator is converted into a separable one by
depolarization. We define \be \tilde\rho=\rho_3 +
\frac{\Delta}{2} (|\Psi_2^+\rangle\langle \Psi_2^+| - |\Psi_2^-\rangle\langle
\Psi_2^-|). \ee This operator is positive since $\Delta \le 2 \lambda_2$
[equivalently, $\rho_3^{T_A}\ge 0$, cf. (\ref{pt})] and fulfills
$\tilde\rho^{T_A}=\tilde\rho$. It has been shown in \cite{Le99}
that all states in $\C^2
\otimes \C^N$ which fulfill $\tilde\rho^{T_A}=\tilde\rho$ are
separable, thus the last property ensures separability of
particle $A$. Furthermore, the state $\tilde\rho$ can be
depolarized to the state $\rho_3$ \cite{Depol}.

(ii) We show that if $\rho_3^{T_A},\rho_3^{T_B},\rho_3^{T_C} \ge
0$ then $\rho_3$ is fully separable (note again that the
opposite is trivially true). Again, the idea is to define an
operator $\hat\rho$ which can be depolarized into the form
$\rho_3$ by using local operations and that is fully separable.
Let $\hat\rho$ be a state of the form (\ref{rho0}) with
coefficients $\hat
\lambda_0^\pm \equiv \lambda_0^\pm$, and $\hat\lambda_k^\pm \equiv \lambda_k \pm
\Delta/2$ ($k=1,2,3$). Clearly, $\hat\rho$ can be depolarized into $\rho_3$
\cite{Depol}. We now rewrite $\hat\rho$ as follows:
\bea
\hat\rho&=&\frac{1}{2}\sum_{k=0}^3
(\hat\lambda^+_k+\hat\lambda^-_k-\Delta)(|\Psi_k^+\rangle\langle \Psi_k^+| +
  |\Psi_k^-\rangle\langle \Psi_k^-|) \nonumber \\
 &+& \Delta \sum_{k=0}^3 |\Psi_k^+\rangle\langle\Psi_k^+|.\label{sepa}
\eea
Since $\rho_3^{T_A},\rho_3^{T_B},\rho_3^{T_C} \ge 0$, all
coefficients in (\ref{sepa}) are positive. The first term in
(\ref{sepa}) can be written as $\sum_{k=0}^3
\frac{(\hat\lambda^+_k+\hat\lambda^-_k-\Delta)}{2}
(|k,0\rangle\langle k,0|+ |(3-k),0\rangle\langle (3-k),0|)$ and
is thus separable. Let us define $|\phi_0\rangle=|+++\rangle$,
$|\phi_1\rangle=|+--\rangle$, $|\phi_2\rangle=|-+-\rangle$, and
$|\phi_3\rangle=|--+\rangle$, where
$|\pm\rangle=(|0\rangle\pm|1\rangle)/\sqrt{2}$. Using this, the
second term in (\ref{sepa}) can now be written as $ \sum_{k=0}^3
|\phi_k\rangle\langle\phi_k|$ and is thus also separable, which
concludes the proof.
\\
{\bf 2.2. Distillability}

We turn now to analyze the distillability properties of $\rho_3$. We
show that we can distill a maximally entangled state
$|\Phi^+\rangle_{\alpha\beta}$ between $\alpha$ and $\beta$ iff both
$\rho_3^{T_\alpha},\rho_3^{T_\beta}$ are not positive. This
automatically means that if all three partial transposes are not
positive, we can distill a GHZ state (since we can distill an entangled
state between $A$ and $B$ and another between $A$ and $C$ and then
connect them to produce a GHZ state \cite{Zu93}). Furthermore, from our
previous analysis on separability we have that if any state belongs to
class 3 the partial transpose with respect to the third particle is
negative (otherwise it would belong to class 5). As mentioned above, if
we have that $\rho^{T_C}$ is not positive but $\rho^{T_A},\rho^{T_B}\ge
0$ and we have maximally entangled states between $A$ and $B$ at our
disposal, then we can activate the entanglement between $ABC$ and create
a GHZ state \cite{note} (see Table I).

In order to prove the statements concerning distillability, we just have
to show that if $\rho_3^{T_B},\rho_3^{T_C}$ are not positive then we can
distill a maximally entangled state between $B$ and $C$. That this condition
is necessary follows from the fact that by local operations one cannot change the
positivity of the partial transpose and therefore if one is able to
distill \cite{Ho99} (which gives rise to non positive partial
transposes) one must start with non positive partial transposes. Let us
consider first that we perform a projection measurement in $A$ on the
state $|+\rangle$; one can easily show that the remaining state of $B$
and $C$ is purificable iff $\Delta/2 > \lambda_1+\lambda_3$ (which
corresponds to having a fidelity $F>1/2$ between the resulting pair). It
may happen that this condition is not satisfy even though $\Delta/2
>\lambda_1,\lambda_3$ [i.e. $\rho_3^{T_B},\rho_3^{T_C}$ are not
positive, cf.\ (\ref{pt})]. In such a case, we can use the following
purification procedure. The idea is to combine $M$ trios in the same
state $\rho_3$, perform a measurement and obtain one trio with the same
form (\ref{rho0}) but in which the new $\Delta$ is exponentially
amplified with respect to $\lambda_{1,3}$. In order to do that, we
proceed as follows: We take $M$ trios, and apply the operator
$P=|00\ldots 00\rangle\langle 00\ldots 00| + |10\ldots 00\rangle\langle
11\ldots 11|$ in all three locations. This corresponds to measuring
a POVM that contains $P$ obtaining the outcome associated to $P$.
The resulting state $P^{\otimes 3}\rho_3^{\otimes M}(P^\dagger)^{\otimes 3}$ has the
first trio in an (unnormalized) state of the form (\ref{rho0}) but with
$\tilde\Delta/2 = (\Delta/2)^M$, and $\tilde\lambda_k=\lambda_k^M$.
Given that $\Delta/2 >\lambda_1,\lambda_3$, for $M$ sufficiently large we
can always have $\tilde \Delta /2 > \tilde \lambda_1+\tilde\lambda_3$.
This implies that if we project $A$ onto $|+\rangle$ in this new trio we
will have a state between $B$ and $C$ with $F>1/2$ and therefore that
can be purified to a maximally entangled state.
\\
{\bf 3. Multi--qubit systems}

We now generalize the above results to the case of $N\ge 3$
qubits. We will just quote the results here, since the
corresponding proofs are similar to those of the three--qubit
case (see \cite{Du99}). We consider the family of states
\bea
\label{rhoN}
\rho_N &=& \sum_{\sigma=\pm} \lambda_0^\sigma |\Psi^\sigma_0\rangle\langle
  \Psi^\sigma_0| \nonumber\\
&& + \sum_{j=1}^{2^{(N-1)}-1} \lambda_j (|\Psi^+_j\rangle\langle \Psi^+_j|
  + |\Psi^-_j\rangle\langle \Psi^-_j|).
\eea
with the GHZ--basis
\be
\label{notation}
|\Psi^\pm_j\rangle \equiv \frac{1}{\sqrt{2}} (|j\rangle|0\rangle
  \pm |(2^{N-1}-j-1)\rangle|1\rangle),
\ee
and $j$ is again understood in binary notation. As before, using spin
flip and phaseshift operations one can depolarize any state of $N$ qubits
into this form\cite{Du99}. We will denote as $A_1,A_2,\ldots,A_N$ the different
qubits. One can readily check that the partial transpose of this
operator with respect to the qubit $A_N$ is positive iff $\Delta\equiv
\lambda_0^+ -\lambda_0^-\le 2\lambda_{2^{N-1}-1}$ and similarly for
the rest of the qubits.
\\
{\bf 3.1. Separability}

On one hand we have that if $\rho_N^{T_{A_k}}\ge 0$ then it can
be written in the form $\rho_N = \sum_i |a_i\rangle_{A_k}\langle
a_i| \otimes |\varphi_i\rangle_{\rm rest}\langle \varphi_i|$,
and therefore $\rho_N$ is separable with respect to particle
$A_k$. On the other hand, if considering all possible partitions
of the qubits in two sets it turns out that for each partition
the partial transpose with respect to one of the sets is
positive then $\rho_N$ is fully separable.
\\
{\bf 3.2. Distillability}

In order to analyze the distillability of a maximally entangled state between
particles $A_i$ and $A_k$ let us consider all possible partitions of the $N$
qubits into two sets such that the particles $A_i$ and $A_k$ belong to different
sets. If for all such partitions, the partial transposition with respect to one
set is negative then distillation is possible.
\\
{\bf 3.3. Example}

Finally, we will apply our results to the case in which we have a
maximally entangled state of $N$ particles mixed with the completely
depolarized state
\be
\rho(x) = x |\Psi_0^+\rangle\langle \Psi_0^+| + \frac{1-x}{2^N} 1.
\ee
These states have been analyzed in the context of robustness of
entanglement \cite{Vi99}, NMR computation \cite{Br99}, and
multiparticle purification \cite{Mu98}. In all
these contexts bounds are given regarding the values of $x$ for which
$\rho(x)$ is separable or purificable. For example, in Refs.\
\cite{Br99,Vi99} they show that in the case $N=3$ if $x\le
1/(3+6\sqrt{2}),1/25$ then the state is separable, respectively. In Ref.
\cite{Mu98} it is shown that for $N=3$ if $x>0.32263$ then $\rho(x)$ is
distillable. Using our results we can state that $\rho(x)$ is fully
non--separable and distillable to a maximally entangled state iff
$x>1/(1+2^{N-1})$, and fully separable otherwise. Specializing this for
the case $N=3$ we obtain that for $x>1/5$ it is non--separable and
distillable.

In summary, we have given a full characterization of the entanglement
and distillability properties of a family of states of $3$ qubits. These
states play the role of Werner states in these systems since any state
can be reduced to such a form by depolarization. Thus, our results
provide sufficient conditions for non--separability and distillability
for general states. In particular, if a state $\rho$ after
depolarization belongs to the class $k$, then $\rho$ must be in a class
$k'\le k$. We have generalized our results to an arbitrary number of
particles.

We thank M. Lewenstein, S. Popescu, G. Vidal and P. Zoller for
discussions. R. T. thanks the University of Innsbruck for hospitality.
This work was supported by the \"Osterreichischer Fonds zur F\"{o}rderung
der wissenschaftlichen Forschung, the European Community under the TMR
network ERB--FMRX--CT96--0087, and the Institute for Quantum Information
GmbH.


\narrowtext
\begin{table}
\begin{tabular}[t]{||c|l|l||}
 Positive Operators         & Class & Distillability                     \\ \hline
 None                       &  1    & (GHZ) $|\Psi_0^+\rangle_{ABC}$     \\ \hline
 $\rho_3^{T_A}$             &  2    & (Pair) $|\Phi^+\rangle_{BC}$       \\ \hline
 $\rho_3^{T_A},\rho_3^{T_B}$&  3    & Activate with $|\Phi^+\rangle_{AB}$\\ \hline
 All                        &  5    &                                    \\
\end{tabular}
\caption[]{Separability and distillability classification of $\rho_3$}
\label{Table1}
\end{table}


\begin{references}


\bibitem{Be65}
J. S. Bell, Physics (N.Y.) {\bf 1}, 195 (1964).

\bibitem{We89}
R. F. Werner, Phys. Rev. A {\bf 40}, 4277 (1989).

\bibitem{Po94}
S. Popescu, Phys. Rev. Lett. {\bf 74}, 2619 (1995).

\bibitem{Wo98}
W. K. Wootters, Phys. Rev. Lett. {\bf 80}, 2245 (1998).


\bibitem{Be96}
C. H. Bennett et al., Phys. Rev. Lett. {\bf 76}, 722 (1996); C. H.
Bennett, et al., Phys. Rev. A {\bf 54}, 3824 (1996).

\bibitem{Pe96}
A. Peres, Phys. Rev. Lett. {\bf 77}, 1413 (1996).

\bibitem{Ho97}
P. Horodecki, Phys. Lett. A {\bf 232}, 333 (1997).

\bibitem{Be93}
C. H. Bennett, {\it et al.}, Phys. Rev. Lett. {\bf 70}, 1895 (1993).

\bibitem{De96}
D. Deutsch, A. Ekert, C. Macchiavello, S. Popescu, and A. Sanpera,
Phys. Rev. Lett. {\bf 77 }, 2818 (1996).

\bibitem{Noisy}
H.-J. Briegel, {\it et al.} Phys. Rev. Lett. {\bf 81}, 5932 (1998); W.
D\"ur {\it et al.}, Phys. Rev. A {\bf 59}, 169 (1999); S. J. van Enk, J.
I. Cirac, and P. Zoller, Phys. Rev. Lett., {\bf 78}, 4293(1997); S. J.
van Enk, J. I. Cirac, and P. Zoller, Science, {\bf 279}, 205 (1998).

\bibitem{PT}
For a definition of the partial transpose see \cite{Pe96}. We
say a state has positive partial transpose ($\rho^{T_A} \geq 0$)
iff all eigenvalues of $\rho^{T_A}$ are non--negative. Similary,
we use the term ``negative partial transpose'' if at least one
eigenvalue is negative.

\bibitem{Ho97b} 
M. Horodecki, P. Horodecki and R. Horodecki, Phys. Rev. Lett. {\bf
78}, 574 (1997). 


\bibitem{Mu98}
M. Murao et al., Phys. Rev. A {\bf 57}, 4075 (1998).

\bibitem{multi}
J. Kempe, quant-ph 9902036;
A. V. Thapliyal, quant-ph 9811091;
G. Vidal, quant-ph 9807077;
N. Linden and S.
Popescu, Fortsch.Phys. {\bf 46}, 567 (1998);
N. Linden, S. Popescu and A. Sudbery, quant-ph 9801076.

\bibitem{Br99}
S. L. Braunstein et al., quant-ph 9811018

\bibitem{Vi99}
G. Vidal and R. Tarrach, Phys. Rev. A {\bf 59}, 141 (1999).

\bibitem{Gr89}
D. M. Greenberger, M. Horne, A. Zeilinger, {\it Bell's theorem,
Quantum Theory, and Conceptions of the Universe,} ed. M. Kafatos,
Kluwer, Dordrecht 69 (1989); D. Bouwmeester et al.,
Phys. Rev. Lett. {\bf 82 }, 1345 (1999).

\bibitem{note2}
We will omit the symbol $\otimes$ when it is clear from the notation.

\bibitem{Be98} C. H. Bennett et al. , quant-ph 9808030.

\bibitem{note}
One can show this by noting that the singlets allow to teleport states
between locations $A$ and $B$. Thus, for all practical purposes we can
consider a pair of qubits $AB$ as a four--level system in which we can
perform arbitrary operations. The situation is equivalent to that
in which one has 2--level systems entangled to 4--level systems such
that the density operator describing one pair acts on $\C^2\otimes \C^4$
and has a negative partial transpose. It can be easily shown [W. D\"ur,
J. I. Cirac, M. Lewenstein and D. Bru\ss, quant-ph 9910022] that in systems $2\times
N$ negative partial transpose is a necessary and sufficient condition
for distillation, and therefore one can distill arbitrary states. Using
again teleportation, one can end up with a GHZ state shared at $A$, $B$
and $C$.

\bibitem{Ho99}
M. Horodecki, P. Horodecki and R. Horodecki, Phys. Rev. Lett. {\bf 82},
1056 (1999).

\bibitem{Depol}
By mixing we understand in the following that a certain
operation is (randomly) performed with $p=\frac{1}{2}$, while
with $p=\frac{1}{2}$ no operation is performed. It is easy to
check that the following sequence of mixing operations is
sufficient to make $\rho$ diagonal in the basis (\ref{GHZbasis})
without changing the diagonal coefficients: In the first round
we apply simultaneous spin flips at all 3 locations; in the
second and third round we apply $\sigma_z$ to systems B and C or
A and C respectively. Finally, one can depolarize the subspaces
spanned by $\{|\Psi_j^{\pm}\rangle\}$ for each $j>0$ by using random
operations that change $|0\rangle_\alpha\to e^{i\phi_\alpha}
|0\rangle_\alpha$ ($\alpha=A,B,C$) with
$\phi_A+\phi_B+\phi_C=2\pi$ (this condition ensures that
$\lambda_0^{\pm}$ remains unchanged).

\bibitem{Le99}
M. Lewenstein, J. I. Cirac and S. Karnas, quant-ph/9903012.

\bibitem{Zu93}
M. Zukowski, {\it et al}, Phys. Rev. Lett. {\bf 71}, 4287 (1993).

\bibitem{Du99}
Details of the proofs are given in W. D\"ur and J. I. Cirac, quant-ph 9911044.

\end{references}
\end{document}